\documentclass[final,5p,times,twocolumn,authoryear]{elsarticle}

\usepackage{amssymb}
\usepackage{xcolor}
\usepackage{amsmath}
\usepackage[utf8]{inputenc}
\usepackage[T1]{fontenc}
\usepackage{tabularx}
\usepackage[pdfencoding=auto, psdextra]{hyperref}

\newcommand{\kms}{km\,s$^{-1}$}

\journal{Physics of the Dark Universe}

\begin{document}

\begin{frontmatter}



\title{Is \texorpdfstring{$\omega_0 \omega_a$}{}CDM a good model for the clumpy Universe?}

\author[label1]{Fernanda Oliveira}
\affiliation[label1]{organization={Observatório Nacional}, 
addressline={Rua General José Cristino, 77, São Cristóvão},
city={Rio de Janeiro},
postcode={20921-400},
state={RJ},
country={Brazil}}

\author[label1]{Felipe Avila}
%\author[first]{}
%\affiliation[first]{organization={Observatório Nacional},%Department and Organization
%addressline={Rua General José Cristino, 77, São Cristóvão},
%city={Rio de Janeiro},
%postcode={20921-400},
%state={RJ},
%country={Brazil}}
\author[label1]{Camila Franco}
\author[label1]{Armando Bernui}

\begin{abstract}
The DESI collaboration just obtained a set of precise BAO measurements, that combined with CMB and SNIa datasets show that 
the $\omega_0 \omega_a$CDM model is preferred over $\Lambda$CDM, at more than $4\,\sigma$, 
to describe the dynamics of the expanding Universe. 
This raises the question whether this model also suitably describes  the clumpy Universe. 
Also lately, detailed analyses of diverse cosmic tracers resulted in a new dataset of measurements of an observable from the clumpy Universe: $\sigma_8(z)$, spanning a high-redshift data $z \in [0.013, 3.8]$. 
In this work we use this dataset of 15 $\sigma_8(z_i)$ measurements to study the viability 
of the $\omega_0 \omega_a$CDM cosmological model to explain the clustered Universe. 
Our analyses compare the $\omega_0 \omega_a$CDM model with the $\sigma_8(z)$ function 
reconstructed from the data points using Gaussian Process. 
Moreover, we perform a similar evaluation of the $\Lambda$CDM model considering Planck 
and~DESI best-fit parameters. 
In addition, we implemented robustness tests regarding Gaussian Process reconstruction to support our results.
\end{abstract}
%Example abstract for the Physics Letters B journal. Here you provide a brief summary of the research and the results.




\end{frontmatter}

%\tableofcontents

\section{Introduction}
\label{introduction}

%--with high statistical significance-- %combined with other data
%DESI (Dark Energy Spectroscopic Instrument) 
Analyses of exceptionally precise measurements of the Baryon Acoustic Oscillations (BAO) phenomenon, recently obtained by the Dark Energy Spectroscopic Instrument (DESI) collaboration~\citep{DESI2013}, 
suggest a turning point in the determination of the standard cosmological model. 
%
%$w$ is the ratio of pressure density to energy density (CPL)
Considering the Chevallier-Polarski-Linder (CPL) parametrization
~\citep{Chevallier2000,Linder2002}
for the time-dependent dark energy equation of state, 
$\omega(a)=\omega_0 + \omega_a\,(1-a)$, 
where $a$ is the space-time scale factor normalized to $a=1$ at present time, 
the DESI analyses combined BAO, CMB, 
and SNIa\footnote{CMB means cosmic microwave background, and SNIa 
means supernovae type Ia} data to find a highly significant preference 
for the $\omega_0 \omega_a$CDM model as compared with the flat-$\Lambda$CDM, 
from now on $\Lambda$CDM, the model with constant dark energy 
$\omega_0=-1,\, \omega_a=0$~\citep{DESI2025}~\footnote{The DESI results appear robust with other parametrizations~\citep{DESI2025b}; for a different point of view see, e.g.,~\cite{Nesseris2025}.}. 
%\textcolor{blue}{CF: BAO aparece por extenso na primeira menção durante o texto, mas CMB, e SNIa, não.}

This result, showing that this model suitably describes the accelerated expansion of the background Universe, poses the question if the $\omega_0 \omega_a$CDM model also explains satisfactorily the data from the clustered Universe~\citep{Mukhanov1992,Coles1996,Brandenberger2011,Alam2021,Avila22a}. 
In fact, competing cosmological models should also correctly describe the clustering properties of the Universe at large scales, as well as the $\Lambda$CDM does~\citep{Marques2020, Avila21, Sahlu2024, Novaes2025, Liu2025, Vanetti2025}. 
Currently, alternative cosmological models, like modified gravity models, $F(R)$, or interaction dark energy (IDE) models, 
are being probed at the perturbative level measuring their capability to reproduce the growth of cosmic structures data, namely $f$ or 
$f\sigma_8$~\citep{Clifton2011,Wang2016,Bernui2023,Colgain2024,Park2024,Toda2024,Liu2025,Pan25,Akarsu2025}. 
Although the data from the perturbed Universe are not as precise as one would like, they are still suitable for testing the viability of alternative 
models~\citep{Basilakos2017,Nesseris2017,Perenon2019,Felegary2024,Sahlu2024,Oliveira2025}.

%the RMS matter fluctuations amplitude
In recent years, there have been efforts to measure, at several redshifts, another important cosmic observable from the clustered Universe: 
$\sigma_{8,0} \equiv \sigma_8(z=0)$ 
%$\sigma_8(z)$, 
the present-day matter fluctuations amplitude at the scale of $8$ Mpc$/h$. 
As a result, currently we have a set of 15 measurements in the redshift interval $z \in [0.013,\,3.80]$~\citep{Garcia2021, Abbott2023, Miyatake2022, Farren2024, Piccirilli2024, Franco2025a}.

Our methodology to investigate if the $\omega_0 \omega_a$CDM model describes suitably the data from the clumped Universe is developed in two steps. 
Firstly, we use Gaussian Processes (GP) to reconstruct in a model-independent way the function $\sigma_8(z)$, termed $\sigma_8^{\text{rec}}(z)$, from this set of 15 $\{ \sigma_8(z_i) \}$ measurements~\citep{Seikel2012, Jesus2020, Valente18, Yang15}. 
After that, we examine if the $\omega_0 \omega_a$CDM model obtained from DESI analyses properly fits the function $\sigma_8^{\text{rec}}(z)$. 
This statistical evaluation includes the comparison of the function 
$\sigma_8^{\text{rec}}(z)$ with the $\Lambda$CDM model obtained considering the best-fit parameters found by the Planck~\citep{Planck2018} and DESI~\citep{DESI2025} analyses.

%evolving dark energy EoS
%Let us note, however, the statistical evidence decreases when parametrizations non-equivalent to CPL are used~\citep{Nesseris2025}. 
%$w0wa$CDM cosmological model 

%%--------------------------------------------------
\section{\texorpdfstring{$\sigma_8$}{} data and 
GP reconstruction}\label{GP-recon}
%\lipsum[1]

We consider the dataset of 14 $\sigma_8(z)$ measurements compiled by~\cite{Piccirilli2024}, plus 1 recent measurement at low-redshift by~\cite{Franco2025a}. 
These 15 $\sigma_8(z)$ data points were obtained analysing different cosmic tracers at redshift range $z \in [0.013,\,3.80]$. 
The list of these data, with their corresponding references, is shown in Table~\ref{table1}. 
To investigate if the $\omega_{0} \omega_{a}$CDM model describes well the clumped Universe our first step is to use GP to reconstruct, in a model-independent way, the function $\sigma_8(z)$ in the interval $z \in [0,\,4]$, function that we shall denote by $\sigma_8^{\text{rec}}(z)$.

%\textcolor{red}{However, caution is needed for the data point $(z,\sigma_8(z)) = (1.10,0.48 \pm 0.01)$. We observed, in preliminary analyses, that this datum has the potential to bias our GP reconstruction due to its very small measurement error. For this, we decided to remove it from our statistical analyses.}

\begin{table} 
\begin{centering}
\begin{tabular}{|l|l|l|l|} \hline
$z$  & $\sigma_{8}(z)$ & error & References \\ \hline
$0.013^{\star}$ & 0.78 & 0.04 & \cite{Franco2025a}\\ \hline
0.24 & 0.67            & 0.04 & \cite{Garcia2021}\\ \hline
0.47 & 0.58            & 0.04 & \cite{Abbott2023}\\ \hline
0.53 & 0.59            & 0.03 & \cite{Garcia2021}\\ \hline
0.60 & 0.59            & 0.02 & \cite{Farren2024}\\ \hline
0.63 & 0.53            & 0.04 & \cite{Abbott2023}\\ \hline
0.69 & 0.66            & 0.10 & \cite{Piccirilli2024}\\ \hline
0.80 & 0.47            & 0.04 & \cite{Abbott2023}\\ \hline
0.83 & 0.58            & 0.04 & \cite{Garcia2021}\\ \hline
0.92 & 0.44            & 0.06 & \cite{Abbott2023}\\ \hline
1.10 & 0.48            & 0.01 & \cite{Farren2024}\\ \hline
1.50 & 0.46            & 0.05 & \cite{Garcia2021}\\ \hline
1.59 & 0.39            & 0.06 & \cite{Piccirilli2024}\\ \hline
2.72 & 0.22            & 0.06 & \cite{Piccirilli2024}\\ \hline
3.80 & 0.12            & 0.06 & \cite{Miyatake2022}\\ \hline
\end{tabular}
\caption{
The $\sigma_8$ datum at $z=0.013^{\star}$ was obtained at the scale of $8$~Mpc. To convert it to Mpc/$h$ units one needs to assume a value for $h$. 
Thus, e.g., the value appearing in the first row of this table was obtained assuming $h = 0.6727$, from~\cite{Planck2018}. 
%$h = 0.6727; 0.6980; 0.7304$ from references~\cite{Planck2018},~\cite{Freedman21}, and~\cite{Riess22}, 
%one obtains: $0.78 \pm 0.04; 0.80 \pm 0.05; 0.83 \pm 0.05$, 
}
\label{table1}
\end{centering}
\end{table}

%We use GP to reconstruct the function $\sigma_8(z)$ from 
%this 15 $\sigma_8(z)$ measurements. 

As a matter of fact, GP have become the main statistical tool for reconstructing cosmological parameters in a non-parametric way. 
%~\citep{Seikel2012,Seikel13,Yang15, Valente18}. 
It allows us to study various problems independently of an underlying cosmological model and have been largely applied in modern cosmology~\citep{Oliveira2023, Seikel2012, Seikel13, Avila22a, Avila22b, Zhan2018, Jesus2020, Mukherjee21, OColgain2021, Mu2023, Dinda2023, 
escamilla2023, LHuillier2019imn, Yang15, Valente18, Avila2025}. 
GP are a generalisation of Gaussian distributions that characterise the properties of functions~\citep{Rasmussen06}. 
They are fully defined by their mean function and covariance function, $m(\textbf{x})$ and $k(\textbf{x},\textbf{x}')$,
\begin{eqnarray}
 m(\textbf{x}) &=& \mathbb{E}[f(\textbf{x})]\,,\nonumber\\
 k(\textbf{x},\textbf{x}') &=& \mathbb{E}[(f(\textbf{x})-m(\textbf{x}))(f(\textbf{x}')-m(\textbf{x}'))]\,, 
\end{eqnarray}
then we write the GP as
\begin{equation}
f(\textbf{x}) \sim \mathcal{GP}(m(\textbf{x}), k(\textbf{x},\textbf{x}'))\,.
\end{equation}
Although it is independent of assuming a fiducial cosmological model, 
the GP procedure needs a specific kernel to reconstruct $f(x)$. 
Recently, studies have been carried out to verify the impact of the kernel on reconstructions~\citep{Hwang23,Zhang23}. 
%In~\ref{appendixB}, we redo our analysis using two kinds of the Mat\'ern kernel and see if our results are dependent of the chosen kernel.

To perform the GP reconstruction, we use the public code GaPP\footnote{\url{https://github.com/JCGoran/GaPP}} developed by \cite{Seikel13} following \cite{Rasmussen06}.

%Following \cite{Rasmussen06}, 
%To perform GP, we use the GaPP\footnote{\url{https://github.com/JCGoran/GaPP}} code, developed by~\cite{Seikel2012}, 
%and the algorithm from~\cite{Rasmussen06}. \textcolor{blue}{CF: Tem repositório desse segundo algoritmo? Se sim, seria bom colocar um footnote para ele como foi feito no da Seikel.}
%\textcolor{red}{Esta pergunta ficou para Felipe e/ou Fernanda}

\begin{figure}[ht]
\centering 
\hspace{-0.4cm}
\includegraphics[width=0.5\textwidth]{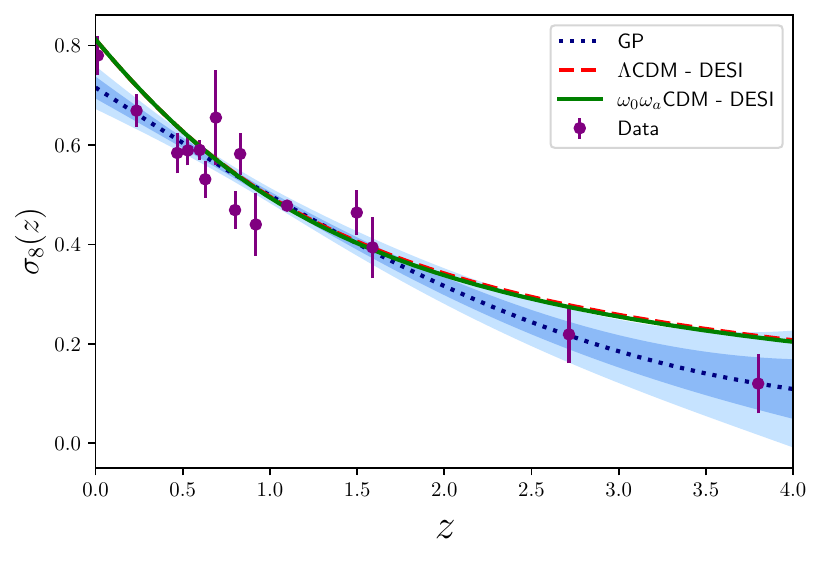}
\caption{Gaussian Process reconstruction for the $\{ \sigma_8(z_i) \}$ 
dataset displayed in Table~\ref{table1}. 
The dotted function represents the GP reconstruction function, $\sigma^{\text{rec}} _8(z)$, with their respective $1\,\sigma$ and $2\,\sigma$ uncertainties represented by dark-blue and light-blue shadows. 
The dashed line (in red) and the continuous line (in green) represent
$\Lambda$CDM and $\omega_{0} \omega_{a}$CDM, respectively. 
For these models we assumed $\sigma_{8}(0) = 0.8120$ from~\cite{Planck2018}. 
}
\label{fig_GP+DESI}%
\end{figure}

%%--------------------------------------

%%------------------------------------------
%A random equation, the Toomre stability criterion:
%\begin{equation}
%Q = \frac{\sigma_v \times \kappa}{\pi \times G \times \Sigma}
%\end{equation}

%\section{Title 3}
%%\label{}
%\lipsum[2]
%\subsection{Subsection title}
%\subsection{Subsection title}
%\lipsum[3]

%%----------------------------------------------
\section{Statistical analyses and Results}
\label{analyses+results}

The evolution of the cosmological observable $\sigma_8(z)$ is described using the linear cosmological perturbation theory, which governs the growth  density fluctuations through the matter density contrast 
$\delta_m(\textbf{r},a)$, with $a(t)$ the scale factor, 
or equivalently $\delta_m(\textbf{r},t)$, dependent on the cosmic time $t$, 
and defined as 
\begin{equation}\label{contrast}
\delta_m(\textbf{r},t) \equiv \frac{\rho_m(\textbf{r},t)-\Bar{\rho}_m(t)}{\Bar{\rho}_m(t)} \,,
\end{equation}
where $\rho_m(\textbf{r},t)$ is the matter density at position 
$\textbf{r}$ and cosmic time $t$, and $\Bar{\rho}_m(t)$ is the background matter density at the same epoch.

In the Newtonian approach, i.e., for sub-horizon scales, one can obtain a second order differential equation to describe the matter fluctuations~\citep{Coles1996,Avila21} 
\begin{equation}\label{eq:edo}
\ddot \delta_m(t) + 2 H(t)\,\dot \delta_m(t) 
- 4 \pi G\, \bar{\rho}_m(t)\,\delta_m(t) = 0 \,,
\end{equation}
where $H(t)\equiv \dot{a}(t)/a(t)$ is the Hubble parameter and $G$ is the Newton gravitational constant.

To solve equation~(\ref{eq:edo}), it is necessary to assume a cosmological model. 
In this work we solve this equation for the $\Lambda$CDM 
and $\omega_0 \omega_a$CDM models, 
where the Hubble parameter is, respectively, 
\begin{equation}\label{hlcdm}
H(a) = H_0 \sqrt{\Omega_{m0} a^{-3} + \Omega_{\Lambda 0}} \,,
\end{equation}
and 
\begin{equation}\label{hubblew0wacdm}
H(a) = H_0 \sqrt{\Omega_{m0}\,a^{-3}  + \Omega_{\Lambda 0}\,a^{ -3 ( 1 + \omega_0 + \omega_a ) }\,e^{-3\omega_a(1-a)} } \,,
\end{equation}
and where $\Omega_{m0}$ and $\Omega_{\Lambda 0}$ are the density parameters of matter and dark energy today, respectively, 
and with the scale factor $a$ related to the redshift $z$ by 
$a = 1 / (1+z)$.

In the linear approximation a solution for 
equation~(\ref{eq:edo}), $\delta_m(z) \sim D(z)$, is given by the growing mode, $D(z)$, equation numerically solved with initial conditions~\citep{Nesseris2017}: 
$\delta(z \gg 1) = 1 / (1+z)$ and $\dot{\delta}(z \gg 1) = 1$.

%in the matter dominated era: $a = 1/(1+z) \sim t^{2/3}$, 

This function $D(z)$ is related to the power spectrum of the density fluctuations $P(k,z)$ by $P(k,z) = D^2(z)P(k,0)$. 
Then, one can compute the variance of matter fluctuations at the scale $R$~\citep{Diemer2018, Franco2025a}
\begin{equation}
\sigma^2(R,z) = \frac{1}{2\pi^2} \int P(k,z) W^2_R (k) k^2 dk \,,
\end{equation}
where $ W^2_R (k)$ is the top-hat window with spherical symmetry.

%power spectrum is usually normalized to give a particular variance

The square root of this     variance at the scale $R = 8 h^{-1}$Mpc 
is the matter fluctuations amplitude, denoted by $\sigma_8(z)$. 
Then, one can write~\citep{Nesseris2017} 
\begin{equation}
\label{eq-sigma8}
\sigma_8(z) = \sigma_{8,0} \,\frac{D(z)}{D(0)} \,. 
\end{equation}
Therefore, given a cosmological model, one numerically solves 
equation~(\ref{eq:edo}) to obtain $D(z)$ and $D(0)$, then uses the equation~(\ref{eq-sigma8}) to compute the function $\sigma_8^{\rm mod}(z)$.
For the results displayed in Table~\ref{table-results}, 
we assumed $\sigma_{8, 0} = 0.8120$ from~\cite{Planck2018}.

According to the analyses described in the previous section we have obtained the GP reconstructed function $\sigma_8^{\text{rec}}(z)$ in the interval $z \in [0,\,4]$, 
which is shown as a dotted curve in Figure~\ref{fig_GP+DESI}. 
Also in this figure we observe the behaviour of the models $\omega_0 \omega_a$CDM and $\Lambda$CDM, both obtained with DESI best-fit 
parameters. 
This illustrative comparison shall be quantified below, but it already shows in advance a similarity of the models in study to fit the data.

We shall perform a statistical comparison between 
$\sigma_{8}^{\text{mod}}(z)$ from a cosmological model in study with respect to the reconstructed function 
$\sigma_8^{\text{rec}}(z)$. 

For this, to measure how well $\sigma_{8}^{\text{mod}}$ fits $\sigma_{8}^{\text{rec}}$, we define 
\begin{equation} \label{chi2red}
\chi^{2}_{\rm mod/rec} \equiv \frac{1}{N}\,\sum_{i=1}^{N} \,\frac{\left[ \sigma_{8}^{\text{mod}}(z_i) - \sigma_{8}^{\text{rec}}(z_i) \right]^{2}}
{\sigma^{\text{mod}}_{\sigma_8}(z_i)^2 + 
 \sigma^{\text{rec}}_{\sigma_8}(z_i)^2 } \,\,,
\end{equation}
where $\textbf{mod}$ and $\textbf{rec}$ refer to the cosmological model in study and to the GP reconstructed function, respectively. 
In these analyses we adopted $N=1000$ bins\footnote{$N$ is the number of bins used in the GP reconstruction. For consistency, we have verified that the final result is independent of 
$N$, for $N > 100$.}. 
Lower values for $\,\chi^2_{\rm mod/rec}$ indicate a better agreement with the 
data\footnote{We are not using the \textit{Akaike Information Criterion}~\citep{Akaike1974} 
because the cosmological parameters of the models studied were regressed using various datasets (e.g., BAO, CMB, SNIa, etc.).}.

Rigorously, we are not adjusting parameters of a given model to fit the data. Instead, we compare the performance of cosmological models to describe the time evolution of the cosmological observable $\sigma_8^{\text{rec}}(z)$. 
For this reason, we use the best-fit parameters of each model including their errors, as reported in the literature~\citep{DESI2025,Planck2018}. 
It is worth noting that the datasets combined by DESI were CMB+BAO+SNIa, 
and their results are summarized in Table V in~\cite{DESI2025}; 
the SNIa dataset they used comes from DESY5~\citep{DES2024}. 
The best-fit cosmological parameters for the models investigated by 
the DESI Collaboration that we are using in our analyses are displayed in our Table~\ref{table-results}.

We notice in Figure~\ref{fig_GP+DESI} that the cosmological models analysed by DESI are competitive in reproducing our GP reconstructed function. 
Using equation~\eqref{chi2red}, we quantify the best-fit analysis of three cosmological models to describe the data from the clustered Universe, where the results of this 
statistical comparison are summarized in Table~\ref{table-results}. 
In fact, regarding the best-fitting of the $\sigma_8^{\text{rec}}(z)$ function, our analyses show that the model $\omega_0 \omega_a$CDM$^{\text{DESI}}$ is preferred over the $\Lambda$CDM$^{\text{Planck}}$ and $\Lambda$CDM$^{\text{DESI}}$ models.

Moreover, we perform a consistency test, suggested by~\cite{Sabogal2024} (see equation (8) and Figure 2 in Section III therein), in which one can compare $\sigma_8^{\text{rec}}(z)$ with $\sigma_8^{\text{mod}}(z)$, obtained for a cosmological model, 
using the definition 
\begin{equation}\label{reldif}
\Delta(z) \equiv \frac{\sigma_8^{\text{rec}}(z) - \sigma_8^{\text{mod}}(z)}{\sigma_8^{\text{mod}}(z)} \,,
\end{equation}
where $\Delta(z)$ is the relative difference between both functions. 
We found that, for the models studied in these analyses, this quantity differs from zero in around $3\,\sigma$ for the redshift interval $z \in [0,4]$, consistent with the results of Table~\ref{table-results}.

Notice that it is important to examine a possible biasing in the Mean Function selection during the GP reconstruction; this analysis is done in ~\ref{appendixA}. 
Additionally, we also test the GP reconstruction procedure for a possible biasing regarding the choice of the kernel used for the reconstruction. 
This study is done in ~\ref{appendixB}.

\begin{table*}
\begin{centering}
%\scalebox{1.1}{
\begin{tabular}{|c||l|l|l|} \hline
Parameters$\backslash$Model & \,$\Lambda$CDM$^{\text{Planck}}$\, & \,$\Lambda$CDM$^{\text{DESI}}$\, & 
\,$\omega_0 \omega_a$CDM$^{\text{DESI}}$\, \\ \hline
$\omega_0$ & $-1$ & $-1$   & $-0.752 \pm 0.057$  \\ \hline
$\omega_a$ &  0  &  0  & $-0.86^{\,+\,0.23}_{\,-\,0.20}$ \\ \hline
$\Omega_m$ & $0.315 \pm 0.0073$ & $0.3027 \pm 0.0036$ & $0.3191 \pm 0.0056$ \\ \hline\hline
$\chi^{2}_{\rm mod/rec}$ & 3.475 & 3.784 & 3.257 \\ \hline
%$\chi^2_{\text{red}}[N=15]$   & ? & ? & ?  \\ \hline
%\textcolor{red}{$\text{AIC}$}           & 56.769 & 62.591 & 61.882 \\ \hline
\end{tabular}
%}
\caption{
Statistical analyses of the $\omega_0 \omega_a$CDM and $\Lambda$CDM models. 
The cosmological parameter in the 2nd column was taken from~\cite{Planck2018}, 
while 
the data in the 3rd and 4th columns come 
from~\cite{DESI2025}. 
Our comparison analyses, displayed in the last row, 
show the performance of these three models to best-fit the 
$\sigma_8^{\text{rec}}$ data. 
Ultimately, the result is: the DESI model 
$\omega_0 \omega_a$CDM$^{\text{DESI}}$ 
fits the $\sigma_8^{\text{rec}}$ data better
than the models $\Lambda$CDM$^{\text{Planck}}$ and $\Lambda$CDM$^{\text{DESI}}$. 
Therefore, the answer to the question in the title is that, 
from the models analysed, the 
$\omega_0 \omega_a$CDM$^{\text{DESI}}$ is indeed 
a competitive model to reproduce the $\{ \sigma_8(z_i) \}$ data
from the clumpy Universe.
}
\label{table-results}
\end{centering}
\end{table*}
%?
%LCDM-Planck: 3113.945137387209
%LCDM - DESI: 3355.4822631138286
%w0waCDM - DESI: 2808.5220659000342
%?
%LCDM - Planck: 3.4745420212686615
%LCDM - DESI: 3.7844746190118075
%w0waCDM - DESI: 2.7488808859698066
%?

%%---------------------------------------------
\section{Summary and Conclusions}
%%\label{}

Model analyses in cosmology are important to be realized both at the background level as well as at the perturbative level, considering the fact that some cosmological models may reproduce well the Universe background but not the clustered Universe, or {\em vice versa}~\citep{Tsujikawa2010,Capozziello2011,Capozziello2011a,Hirano2015,Bessa2021,Perivolaropoulos2022,Ribeiro23}.

In this work, we performed statistical analyses for model comparisons using a new dataset of 15 measurements 
$\{ \sigma_8(z_i) \}$~\citep{Piccirilli2024, Franco2025a}. 
Specifically, our main objective was to study the viability of the $\omega_0 \omega_a$CDM$^{\text{DESI}}$ model 
--the preferred model that explains the dynamics of the Universe according to DESI analyses~\citep{DESI2025}-- to account for measurements from the clustered Universe~\citep{Coles1996,Alam2021,Marques2024,
Toda2024,Franco2025b,Novaes2025}.

In addition, we have performed a consistency test for the models 
studied, finding that the relative difference 
$\Delta(z)$ defined in equation~(\ref{reldif}) between 
$\sigma_8^{\text{rec}}(z)$ and $\sigma_8^{\text{mod}}(z)$ is around $3\,\sigma$, consistent with the results displayed in Table~\ref{table-results} obtained with the estimator $\chi^{2}_{\rm mod/rec}$.

%equation~(\ref{chi2red})
%\textcolor{blue}{for the $\omega_0 \omega_a$CDM$^{\text{DESI}}$ (FO: não deveríamos especificar isso?)}

The final results of our comparison analyses are displayed in the last row of Table~\ref{table-results}, where we quantify the performance of three models to best-fit the $\sigma_8^{\text{rec}}$ data. 
Our conclusion is: the DESI model $\omega_0 \omega_a$CDM$^{\text{DESI}}$ fits the $\sigma_8^{\text{rec}}$ data better than the models $\Lambda$CDM$^{\text{Planck}}$ and $\Lambda$CDM$^{\text{DESI}}$. 
Therefore, the answer to the question in the title is that, from the three models analysed, the $\omega_0 \omega_a$CDM$^{\text{DESI}}$ is indeed a competitive model to reproduce the $\{ \sigma_8(z_i) \}$ data from the clumpy Universe.

%suggested by~\cite{Sabogal2024} (see equation (8) and Figure 2 in Section III therein), and found that the relative difference between $\sigma_8^{\text{rec}}(z)$ and $\sigma_8^{\text{DESI}}(z)$ from the $\omega_0 \omega_a$CDM$^{\text{DESI}}$ model is between $2 - 3\,\sigma$, in good agreement with the result of Table~\ref{table-results}. 

%%---------------------------------------------
\section*{Acknowledgements}
\noindent
FO and CF thank the Coordenação de Aperfeiçoamento de Pessoal de Nível Superior (CAPES) 
for their corresponding fellowships. 
%CF thanks CAPES for the financial support. 
FA thanks to Fundação de Amparo à Pesquisa do Estado do Rio de Janeiro (FAPERJ), Processo SEI-260003/001221/2025, for the financial support. 
AB acknowledges the Conselho Nacional de Desenvolvimento Científico e Tecnológico (CNPq) for the fellowship.

\appendix

%%---------------------------------
\section{Testing the Mean Function selection}
\label{appendixA}

A recurring question about the performance of Gaussian Processes concerns the impact of the choice of the mean function in the GP reconstruction. 
For this, and following~\cite{Oliveira2023}, in this appendix we perform an important test to evaluate the capability of GP procedure in reconstructing the $\sigma_{8}(z)$ function, using observational data and adopting a zero-mean prior. 
The fiducial cosmology was adopted from~\cite{Planck2018}, with the parameter values listed in Table~\ref{tab:input-cosmo}. 

\begin{table}
\centering
\caption{Cosmological parameters from~\citet{Planck2018}.}
\begin{tabularx}{0.8\linewidth}{>{\centering\arraybackslash}X}
\hline
\hline
Cosmological parameters\\
\hline
$\Omega_b = 0.0494$\\
$\Omega_c = 0.2656$\\
$h = 0.6727$\\
$n_s = 0.9649$\\
$\sigma_{8,0} = 0.8120$\\
\hline
\end{tabularx}
\label{tab:input-cosmo}
\end{table}

%theoretical 

The matter fluctuations amplitude for the fiducial cosmology, $\sigma_{8}^{\rm fid}(z)$, is given by 
\begin{equation}
\label{eq:sigma8}
\sigma_{8}^{\rm fid}(z) = \sigma_{8, 0}\,\frac{D(z)}{D(0)}\;,
\end{equation}
where the growing mode function $D(z) = D(z)^{\rm fid}$ was 
obtained assuming the $\Lambda$CDM as the fiducial model, 
equation~(\ref{hlcdm}), then solving the equation~(\ref{eq:edo}) with initial conditions as in~\cite{Nesseris2017}. 
The parameters for the fiducial cosmology, from the Planck Collaboration~\citep{Planck2018}, are shown in Table~\ref{tab:input-cosmo}, and we use the numerical code Core Cosmology Library\footnote{\url{https://ccl.readthedocs.io/en/latest/}}~\citep[\textsc{CCL};][]{Chisari2019}.

The GP reconstruction was performed using the Gaussian Processes in Python code~\citep[\textsc{GaPP};][]{Seikel2012} to reconstruct $\sigma_{8}(z)$ from observational data, using a squared exponential (SE) kernel with hyperparameters $\theta = [\sigma_{f}, l] = [0.5, 2]$. The reconstruction was performed over the redshift range $z \in [0, 4]$ with $200$ points.

To evaluate the robustness of the GP reconstruction with a mean function, we generated $700$ Monte Carlo realizations of the $\sigma_{8}(z)$ data. 
For each realization Gaussian noise was added to the fiducial $\sigma_{8}^{\rm fid}(z)$, with standard deviation corresponding to the observational uncertainties. 
Then, we reconstructed $\sigma_{8}^{\rm rec}(z)$ using GP with zero mean, and the residuals (the difference between the reconstructed and fiducial matter fluctuations amplitude, $\Delta\sigma_{8}(z) = \sigma_{8}^{\rm rec}(z) - \sigma_{8}^{\rm fid}(z)$) was computed. 
The result is the mean of the $700$ realizations. 
% The GP was then applied to each new dataset, 

\begin{figure}
    \begin{minipage}[b]{\linewidth}
    \centering
    \includegraphics[width=\textwidth]{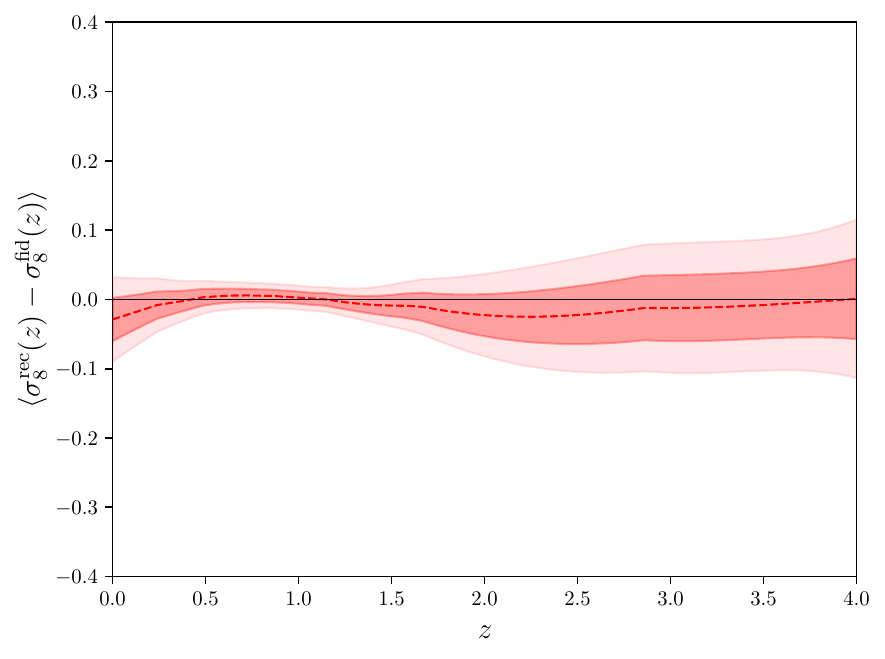}
    \end{minipage}
\caption{Residuals of the GP reconstruction of the function $\sigma_{8}(z)$ using a zero-mean function, computed as $\Delta\sigma_{8}(z) = \sigma_{8}^{\rm rec}(z) - \sigma_{8}^{\rm fid}(z)$. 
The dashed red line represents the mean of $700$ Monte Carlo realizations, while the shaded regions indicate the uncertainties 
within $1\sigma$ and $2\sigma$ confidence levels.}
\label{fig:mean_func}
\end{figure}

As can be seen in Figure~\ref{fig:mean_func}, the difference $\langle\sigma_{8}^{\rm rec}(z) - \sigma_{8}^{\rm fid}(z)\rangle$ is statistically consistent with zero throughout the redshift interval of the reconstruction. 
Therefore, in this analysis, a zero mean function does not introduce significant bias in 
the reconstructed function $\sigma_{8}^{\rm rec}(z)$, provided the underlying function is 
smooth and well-sampled, as is the current case (see the sections~\ref{GP-recon} 
and~\ref{analyses+results}). 
This supports the use of GP for cosmological parameter inference without \textit{ad hoc} 
assumptions about the mean behaviour.

%\label{sec:mean_func}

%As can been see in Figure~\ref{fig:mean_func}, the difference $\langle\sigma_{8}^{\rm rec}(z) - \sigma_{8}^{\rm fid}(z)\rangle$ oscillates near zero across the entire redshift range. 
%The shaded regions confirm that the GP reconstructions are statistically consistent with the fiducial model. 

%Our analyses show that a zero mean function does not introduce significant bias in the reconstruction of $\sigma_{8}(z)$, i.e., $\sigma_{8}^{\rm rec}(z)$, provided the underlying function is smooth and well-sampled, as is the current case (see the 
%section~\ref{GP-recon}). 
%This supports the use of GP for cosmological parameter inference without \textit{ad hoc} assumptions about the mean behaviour.

%%---------------------------------------------------
\section{Kernel test}
\label{appendixB}

We presented our result for the reconstruction of $\sigma_8(z)$ using the SE Kernel, but there are many other kernels available. In this Appendix, we show a comparison between three kernels: SE and Matérn (7/2), SE and Matérn (9/2), and Matérn (7/2) and Matérn (9/2). The Matérn kernel is given by
\begin{equation}
    \label{matern}
    K_{M_{\nu}}(\tau) = \sigma_f ^2 \frac{2^{1-\nu}}{\Gamma(\nu)} \left(\frac{\sqrt{2 \nu} \tau}{l} \right)^\nu K_\nu \left (\frac{\sqrt{2 \nu} \tau}{l} \right),
\end{equation}
where $\Gamma(\nu)$ is the standard Gamma function, $K_\nu$ is the modified Bessel function of second kind and $\nu$ is a strictly positive parameter. $\sigma_f$ and $l$ are hyper-parameters, optimised during the fitting. Since the Matérn kernel tends to SE as $\nu \rightarrow \infty$, Matérn is a kernel that is more sensitive to data fluctuations.

\begin{figure*}
\centering 
\includegraphics[width=0.31\textwidth]{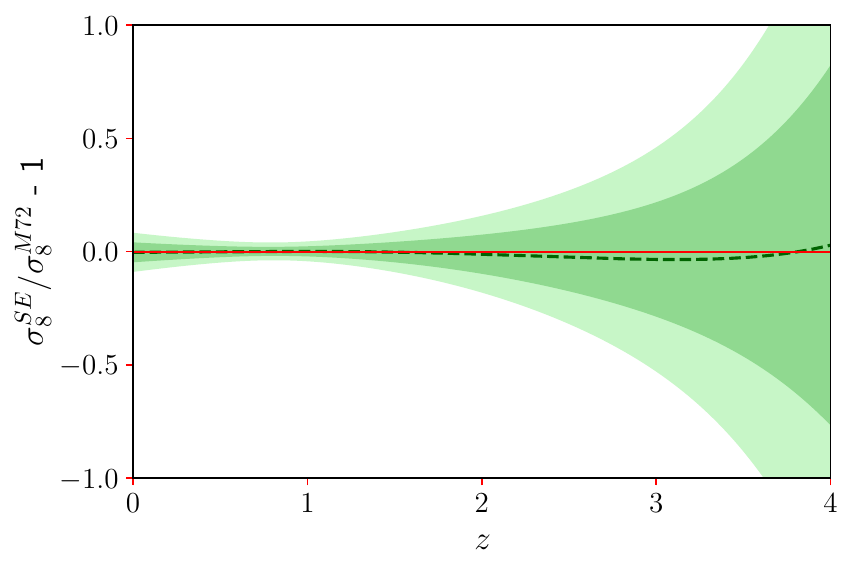}
\includegraphics[width=0.31\textwidth]{
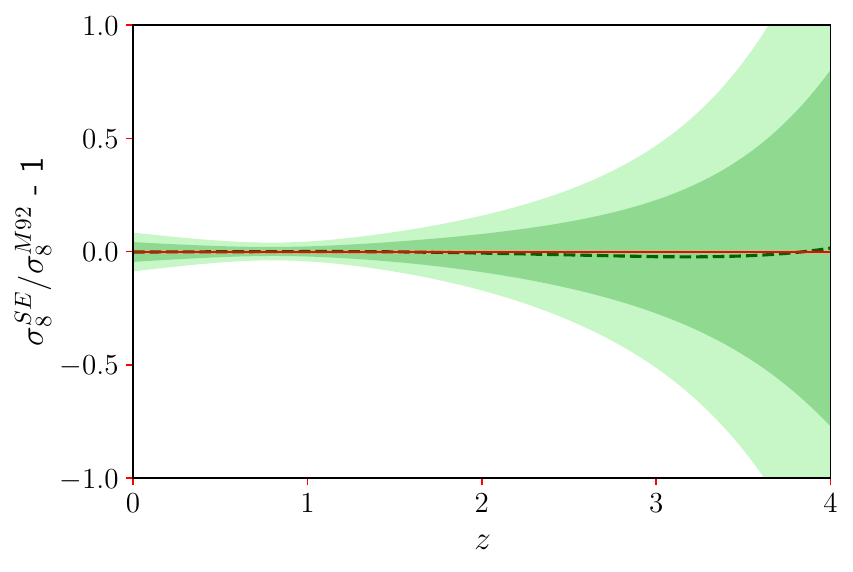}
\includegraphics[width=0.31\textwidth]{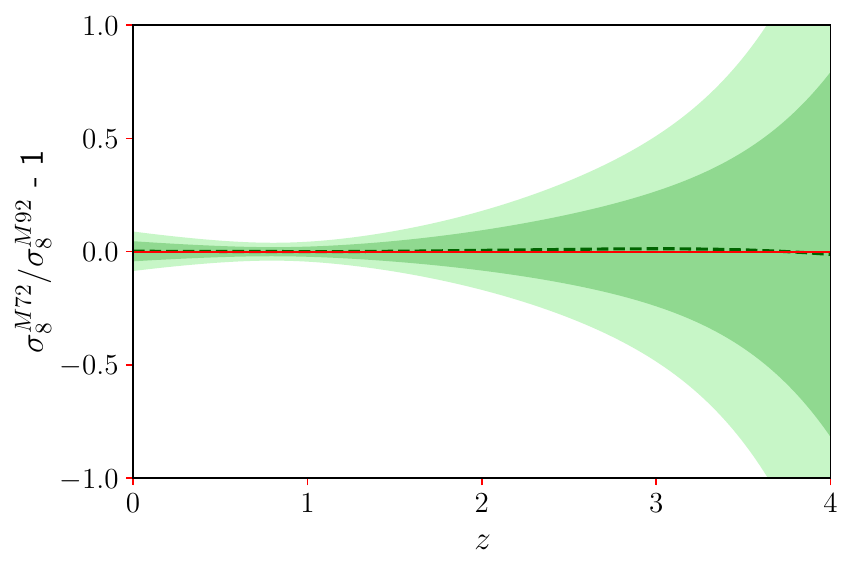}
\caption{Robustness analyses for different kernel reconstructions. \textbf{Left Panel:} Comparison between SE and Matérn (7/2) kernels for the $\sigma_8(z)$ reconstruction. \textbf{Middle Panel:} Same as in the left panel, but between SE and Matérn (9/2). \textbf{Right Panel:} Same as in the left and middle panels, but between Matérn (7/2) and Matérn (9/2). The relative difference obtained with the three kernels is very close to zero in all cases.} 
\label{fig_robust}%
\end{figure*}

In Figure~\ref{fig_robust}, we plot the relative difference between $\sigma^{\text{rec}} _8(z)$ obtained with the three kernels. The relative difference $\mathcal{R}$ is obtained using the equation 
\begin{equation}
\mathcal{R} \equiv 
\frac{\sigma_8^{\rm kernel \,1}}{\sigma_8^{\rm kernel \,2}} - 1, 
\end{equation}
where $\sigma_8^{{\rm kernel}\,\, n}$ is the GP reconstruction curve for the given kernel $n$. 
The errors are estimated by propagating the uncertainties of the GP reconstruction functions, obtaining 
\begin{equation}
\sigma_{\mathcal{R}} = \sqrt{\left( \frac{\partial \mathcal{R}}{\partial \sigma_8^{\rm kernel \,1}} \right)^2 \sigma^2_{\sigma_8^{\rm kernel \,1}} + \left( \frac{\partial \mathcal{R}}{\partial \sigma_8^{\rm kernel \,2}} \right)^2 \sigma^2_{\sigma_8^{\rm kernel \,2}}} \,\,,
\end{equation}
where $\sigma_{\sigma_8^{{\rm kernel} \,\,n}}$ is the uncertainty of the   $\sigma_8^{{\rm kernel} \,\,n}$ reconstruction. 
This test was also applied in~\cite{Oliveira2023} where, as obtained in the analyses presented in this work, the relative difference is close to zero in all cases. 

The choice of the kernel should not mean a big influence on the reconstruction, so we expect that the relative difference should be close to zero. The green shaded areas represent the 1$\sigma$ and 2$\sigma$ CL regions. In all three plots, the red solid line represents the expected value.
As expected, the relative difference is close to zero in all cases until $z = 4$. 
% After this point, the relative difference tends to disagree with the expected value. It can be explained by the lack of $\sigma_8(z)$ data after this redshift.

\bibliographystyle{elsarticle-harv} 
%\bibliography{example}
\bibliography{main_PLB}



%% For example:

%% \bibitem[Aladro et al.(2015)]{Aladro15} Aladro, R., Martín, S., Riquelme, D., et al. 2015, \aas, 579, A101

%%\end{thebibliography}

\end{document}